\documentclass[preprint]{aastex}

\usepackage{epsfig}
\usepackage{emulateapj5}
\newcommand\etal{et~al.}

\newcommand\phots{\:{\rm photons\:cm^{-2}\:s^{-1}}}

\newcommand\LA{Lyman-$\alpha$}

\newcommand\Hone{H\,{\sc i}}

\newcommand\Oone{O\,{\sc i}}

\newcommand\lam{$\lambda$}
\def\deg{\hbox{$^\circ$}}

\slugcomment{Accepted for publication in the {\it Astrophysical Journal}}

\shorttitle{ULTRAVIOLET IMAGING OF POLAR AURORA ON GANYMEDE}
\shortauthors{FELDMAN ET AL.}

\begin{document}

\title{HST/STIS Ultraviolet Imaging of Polar Aurora on Ganymede}

\author{Paul D. Feldman,\altaffilmark{1}
Melissa A. McGrath,\altaffilmark{2}
Darrell F. Strobel,\altaffilmark{1,3} 
H. Warren Moos,\altaffilmark{1}\\
Kurt D. Retherford,\altaffilmark{1}
and Brian C. Wolven\altaffilmark{1}}

\altaffiltext{1}{Department of Physics and Astronomy, The Johns Hopkins 
University, Baltimore, Maryland}

\altaffiltext{2}{Space Telescope Science Institute, Baltimore, Maryland}

\altaffiltext{3}{Department of Earth and Planetary Sciences, The Johns Hopkins 
University, Baltimore, Maryland}


\begin{abstract}

We report new observations of the spectrum of Ganymede in the spectral
range 1160 -- 1720 \AA\ made with the Space Telescope Imaging
Spectrograph (STIS) on HST on 1998 October 30.  The observations were
undertaken to locate the regions of the atomic oxygen emissions at 1304
and 1356 \AA, previously observed with the Goddard High Resolution
Spectrograph on HST, that \citet{hal98} claimed indicated the presence
of polar aurorae on Ganymede.  The use of the $2''$ wide STIS slit,
slightly wider than the disk diameter of Ganymede, produced objective
spectra with images of the two oxygen emissions clearly separated.
The \Oone\ emissions appear in both hemispheres, at latitudes above
$|40|$\deg, in accordance with recent Galileo magnetometer data that
indicate the presence of an intrinsic magnetic field such that Jovian
magnetic field lines are linked to the surface of Ganymede only at high
latitudes.  Both the brightness and relative north-south intensity of
the emissions varied considerably over the four contiguous orbits (5.5
hours) of observation, presumably due to the changing Jovian plasma
environment at Ganymede.  However, the observed longitudinal
non-uniformity in the emission brightness at high latitudes,
particularly in the southern hemisphere, and the lack of pronounced
limb brightening near the poles are difficult to understand with
current models.  In addition to observed solar \Hone\ Lyman-$\alpha$
reflected from the disk, extended Lyman-$\alpha$ emission resonantly
scattered from a hydrogen exosphere is detected out to beyond two
Ganymede radii from the limb, and its brightness is consistent with the
Galileo UVS measurements of \citet{bar97}.

\end{abstract}


\section{Introduction}

Recent observations from both the Galileo spacecraft and the Hubble
Space Telescope (HST) have considerably altered our knowledge of the
atmospheres of the Jovian satellites Europa and Ganymede. Both are now
known to have tenuous atmospheres (column density
$\sim5\times10^{14}$\,cm$^{-2}$) with both molecular oxygen (Europa and
Ganymede) and atomic hydrogen (Ganymede) components.  The Galileo UV
spectrometer detected \Hone\ \LA\ emission at Ganymede from a hydrogen
exosphere \citep{bar97}, and charged particle measurements indicated
that there is also an outflow of protons, implying ongoing gas
production \citep{fra97}.  The oxygen component was detected through
HST/Goddard High Resolution Spectrograph (GHRS) observations of the
atomic oxygen multiplets \Oone\ \lam1304 and \Oone]\,\lam1356
\citep{hal95,hal98}. The intensity ratio of these emissions implies
that the primary source is electron dissociative excitation of
molecular oxygen. The source of both the hydrogen and O$_2$ is thought
to be sputtering of surface water ice by Io plasma torus ions. Beyond
the fact that they exist, very little is known about these atmospheres,
including their vertical structure, areal coverage, and variability,
which could be significant if the dominant source is surface sputtering
because of the asymmetric nature of the plasma bombardment.  Plasma
bombardment of the surface is also supported by the recent detection of
ozone and O$_2$ embedded in the surface ice of Ganymede, and SO$_2$
embedded in the surface ice of Europa \citep{spe95,cal96,nol95,nol96}.

Galileo magnetometer measurements have also shown strong perturbations
in the Jovian magnetic field near Ganymede \citep{kiv96,kiv97}.  The
measured perturbations indicate that the satellite possesses a magnetic
field sufficiently strong ($\sim$1500\,nT) to overpower Jupiter's
ambient field, and that Ganymede's magnetic and spin axes are roughly
aligned \citep{kiv96}.  Near Ganymede closest approach the plasma wave
experiment also detected a significant population of trapped, charged
particles \citep{gur96}, implying that Ganymede possesses a
``magnetosphere within a magnetosphere." Consistent with these results,
the HST/GHRS observations \citep{hal98} have raised the intriguing
possibility that Ganymede exhibits polar aurora. In those spectra the
Ganymede \Oone]\,\lam1356 emission line exhibits a doubly-peaked
profile that is inconsistent with that of a diffuse source filling the
aperture, or with emission from a uniform disk. \citeauthor{hal98}
postulated that the double-peaked profile implies the existence of a
similarly double-peaked structure in the spatial distribution of the
emission source within the aperture, with the strongest emissions
coincident with Ganymede's north and south polar regions.  In this
paper, we report ultraviolet objective grating images of Ganymede made
with the Space Telescope Imaging Spectrograph (STIS) (installed in HST
in February 1997) which confirm the spatial distribution inferred from
the earlier observation and raise new questions about the interaction
of Ganymede's atmosphere with the Jovian magnetosphere.

\section{Observations}

\begin{figure}
\begin{center}
\epsfig{file=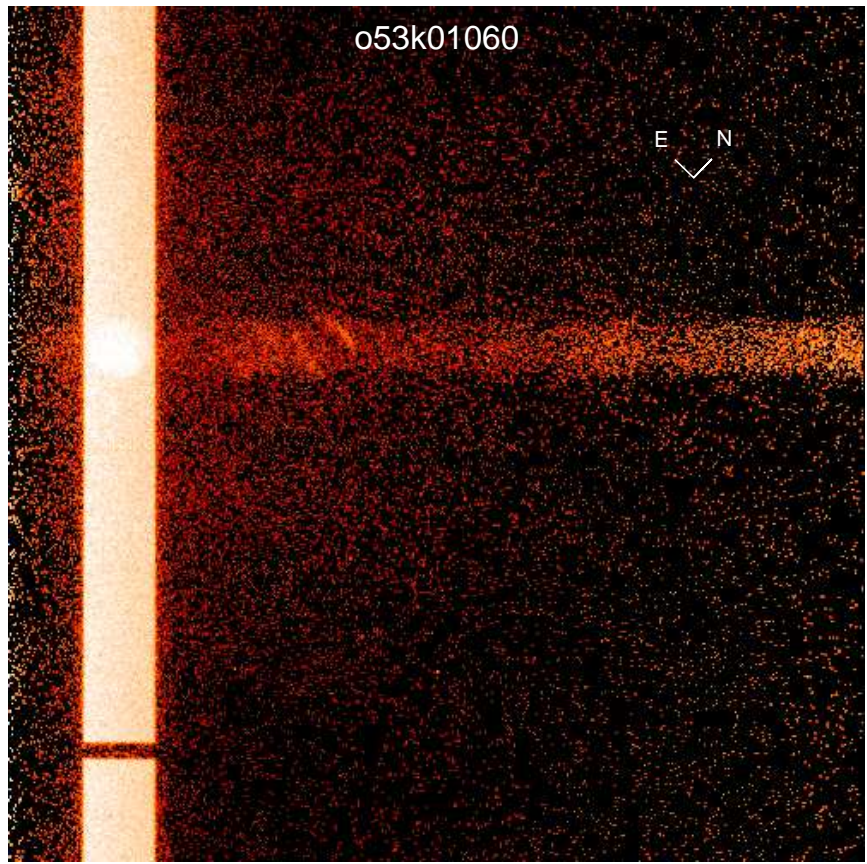, width=4.5in}
\figcaption[]{An example of a single spectral image (ID o53k01060). 
The horizontal axis is wavelength, extending from 1150 to 1720 \AA, 
while the height of the image is $25''$.  Jovian north is indicated.
The vertical stripe on the left is geocoronal \LA\ emission. \label{rawimage}}
\end{center}
\end{figure}

Observations were obtained over four contiguous HST orbits on 1998
October 30 with the STIS G140L grating using the $2'' \times 25''$ slit.  At
the time of observation, Ganymede was 4.25 AU from Earth, its sub-Earth
longitude varied from 290 to 300\deg\ and the phase angle was 8.6\deg.
Since the diameter of Ganymede's disk was $1.\!''71$, this provided
effective objective grating spectroscopy over the wavelength range of
1160 -- 1720 \AA.  A log of the exposures is given in
Table~\ref{tab1}.  The STIS mode used for the observations is the same
as that used to observe Io, and the details can be found in
\citet{roe99}.  Two distinct time-tagged spectral images are obtained
in each of four HST orbits.  The first image in orbits 2--4 exhibits a
very high geocoronal \LA\ background (typically 15 kR as opposed to 3.5
kR during the dark part of an HST orbit) as well as strong
\Oone\ \lam1304 airglow emission due to the illumination of the Earth's
upper atmosphere by sunlight.  An example of a raw spectral image is
shown in Figure~\ref{rawimage}.  Note the $2''$ wide vertical stripe
corresponding to geocoronal \LA\ with the disk reflected \LA\ image of
Ganymede perfectly centered in in the slit.  The raw image also shows
the two oxygen multiplets, clearly separated, and reflected sunlight at
the longer wavelengths.
\placetable{tab1}
\placefigure{rawimage}

\section{Discussion}

\subsection{Spectra}

\begin{figure}
\begin{center}
\epsfig{file=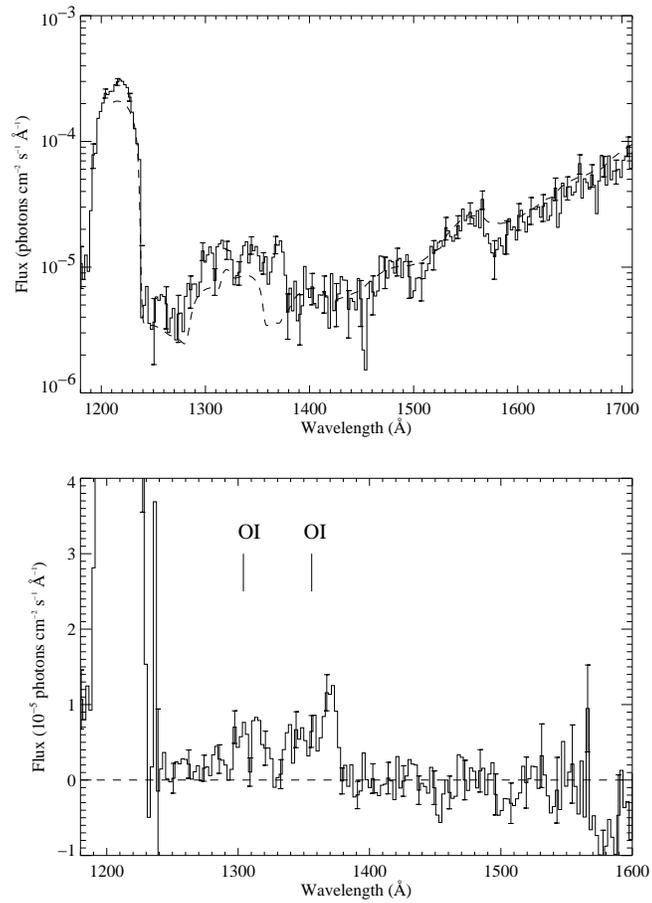, width=3.5in}
\figcaption[]{Top: Extracted spectrum obtained by summing over the
image of Ganymede in Fig.~\ref{rawimage}.  A solar spectrum, convolved with
a uniform disk of Ganymede's diameter, is shown as the dashed curve.  Bottom:
Difference spectrum obtained by subtracting the fitted solar spectrum.  The
positions of the \Oone\ emissions at the center of the disk are indicated.
\label{spect}}
\end{center}
\end{figure}

Individual spectra (two per orbit) were extracted by summing the data
over 82 pixels ($2''$) along the slit centered on Ganymede.  The
background, particularly the geocoronal \LA, was obtained by summing
151 pixels ($3.\!''68$) along the slit on both sides of Ganymede, and
averaging the two.  The results for a single spectral image, with the
background subtracted, are shown in Figure~\ref{spect}.  In this
figure, the one-dimensional extracted spectrum has also been
rebinned by four pixels to enhance the signal/noise ratio.  The unusual
shape of the spectral lines of \Hone\ \LA\ and the two \Oone\ emissions
result from the spatial distribution of these emissions on the disk of
Ganymede.
\placefigure{spect}

Longward of 1380 \AA\ the signal is reflected solar radiation.  To
model this component, a solar spectrum taken with the SOLSTICE
instrument on UARS \citep{woo96} appropriate to the level of solar
activity in October 1998 was convolved with an assumed uniform
reflecting disk of Ganymede's radius (note that one pixel is
$0.\!''0244 \times 0.\!''0244$ and the dispersion is 0.584
\AA\ pixel$^{-1}$).  This is overplotted in Fig.~\ref{spect}.  From
this fit, the planetary albedo can be derived from the ratio of the
reflected flux to the solar flux at Jupiter:

\[ p(\lambda) = \frac{F_{G}(\lambda)\pi d^2}{F_{\odot}(\lambda) 
\Omega_{G}}\phi(\theta,\lambda)   \]

\noindent
where $d$ is the Sun-Jupiter distance (in AU), $\Omega_{G}$ is the solid
angle of Ganymede as seen from Earth, $F_{\odot}(\lambda)$ is the solar
flux at 1 AU, and $\phi(\theta,\lambda)$ is the phase function at phase
angle $\theta$.  To determine the albedo at a wavelength as close as
possible to the oxygen emissions, we choose a 50 \AA\ band centered at
1405 \AA.  With the assumption of unity for $\phi(\theta,\lambda)$, the
derived albedo near 1400 \AA\ is found to be $2.3 \pm 0.2$\%, in good
agreement with the value of $2.6 \pm 0.3$\% derived by \citet{hal98}
from GHRS measurements of the reflected C\,{\sc ii} $\lambda$1335
multiplet.  Note that the measurement of \citeauthor{hal98} was made at a
slightly smaller phase angle, 2.7\deg.

The lower panel of Fig.~\ref{spect} shows a single spectrum after subtraction
of the fitted solar spectrum assuming a constant albedo with wavelength.
The two \Oone\ emissions are clearly separated and both have a shape
determined by the spatial distribution on the disk, similar to that
inferred by \citet{hal98} from their one-dimensional spectra.  The fluxes
of the two multiplets are extracted and tabulated, together with the ratio
of the two (\Oone]\,\lam1356/\Oone\ \lam1304), in Table~\ref{tab1}.  The
values are generally consistent with those reported by \citeauthor{hal98}
(note that \citeauthor{hal98} give the total \Oone\ \lam1304 flux, airglow
plus reflected solar radiation), but the orbit to orbit variation suggests
a real variability, one that is correlated with the changes in the morphology
of the emissions discussed in the next section.  The ratio of \Oone]\,\lam1356 
to \Oone\ \lam1304 is, like that found by \citeauthor{hal98}, consistently
lower (but within the $3\sigma$ uncertainty) than the values expected for
electron impact excitation of O$_2$ alone, 1.6--2.0, indicating a possible
contribution from electron impact of atomic oxygen to the \Oone\ \lam1304
emission.  However, this is not quantifiable as there is an indication,
from Fig.~\ref{spect}, that the albedo is increasing below 1300 \AA\
in which case the \Oone\ \lam1304 flux would be over-estimated and the
true intensity ratio would be closer to the known value for O$_2$ excitation.

\subsection{\Oone]\,\lam1356 Images}

\begin{figure}
\begin{center}
\epsfig{file=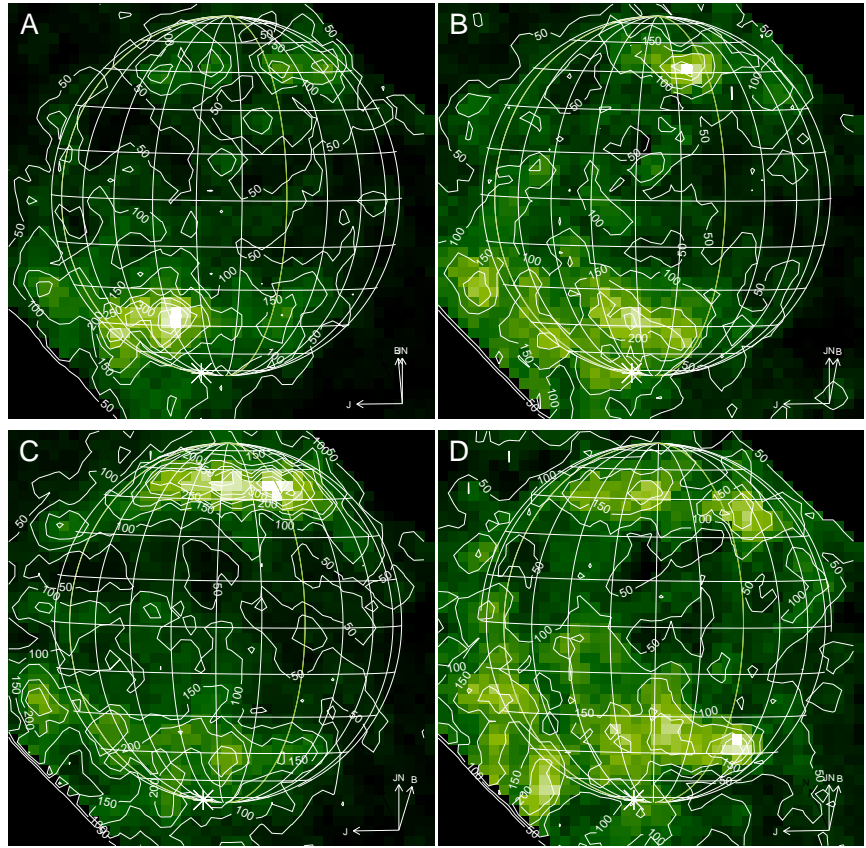, width=4.5in}
\figcaption{Images of \Oone]\,\lam1356 emission for each orbit (indicated
by A--D) are shown, together with brightness contours in rayleighs.  The
compass shows Jovian north (JN), the direction to Jupiter (J) and the
anti-direction of the Jovian magnetic field (B).  \label{oimage}}
\end{center}
\end{figure}

Images of \Oone]\,\lam1356 were constructed using the flat-fielded
counts (``flt'') files from the HST pipeline rather than the fluxed
two-dimensional image (``x2d'') files used to generate
Fig.~\ref{rawimage}.  This was done to avoid distortion introduced by
the changing sensitivity across the $2''$ wide slit, which spans 47
\AA.  To allow for temporal variability, the two separate exposures from each 
orbit were added together since the \Oone]\,\lam1356 emission is not
affected by the higher \LA\ or \Oone\ \lam1304 background levels.
Detector background was evaluated away from Ganymede along the slit and
subtracted from the resulting $82 \times 82$ pixel array.  Each image
was then rotated to align Jovian north along the vertical axis,
rebinned to $41 \times 41$ pixels (each pixel now $0.\!''049$ on a
side), and smoothed by 3 in both directions.  The results are shown in
Figure~\ref{oimage}, together with brightness contours calibrated in rayleighs.  The images are characterized by bright polar regions and appear to be variable 
with time.
\placefigure{oimage}

The location of the emissions, mostly at geographic latitudes above $|40|$\deg\ 
in both hemispheres, is in agreement with the model of \citet{kiv97} that
predicts the presence of Jovian field lines linked to Ganymede only at high
latitudes.  In addition, the regions of brightest emission occur at the
geomagnetic latitudes where the separatrix regions intersect the atmosphere
and which define the boundaries of the polar caps \citep{neu98}.  Magnetic
field line reconnection occurs along the separatrix imparting an
atmospheric signature of enhanced conductivity and current known as the
auroral or polar electrojets.  Over the course of the four orbits the
orientation of the Jovian magnetic field relative to Ganymede's magnetic
field, which is tilted 10\deg\ from the spin axis \citep{kiv97}, varied
considerably, as indicated in Table~\ref{tab1}.  Thus, the locations of the
polar caps and separatrix regions on the surface of Ganymede vary
considerably over a Jovian rotation.  This can account for both the
variations in total \Oone\ flux and the relative brightnesses of the
northern and southern hemisphere emissions.  

However, there are two surprising aspects to these images, because Jovian
magnetospheric electrons have direct access to the polar cap atmospheres,
as implied by particle and fields measurements on Galileo
\citep{wil98,par99}.  They are the longitudinal non-uniformity in the
emission brightness at high latitudes, particularly in the southern
hemisphere, and the lack of pronounced limb brightening near the poles,
even at the smoothed 200 km spatial resolution of our images.  The images
yield $\sim 50$ R limb intensity above the polar caps. Under uniform
conditions this is equivalent to $\sim 15$ R at 60\deg\ latitude and below
our detection limit.  The brightest regions on the disk are 20 times the
disk intensity constrained by the observed limb intensity.

To explore the significance of the observed oxygen brightness, a model
atmosphere was constructed with surface O$_2$ density of $1 \times 10^8$
cm$^{-3}$ and column density of $5.2 \times 10^{14}$ cm$^{-2}$, which is
consistent with the range adopted by \citet{hal98} and at the
abundance upper limit deduced from a UV stellar occultation observed by
Voyager \citep{bro81}.  For the electron density, a model was generated
from measurements by the Galileo plasma wave instrument along fly-by
trajectories and extrapolated to the surface with density of 370 cm$^{-3}$
\citep{gur96}.  Unfortunately, there are no observational constraints on the
electron temperature.  From the data of \citet{sit87} at their maximum
L shell of 13, one would expect the Jovian
magnetospheric electron temperature to be at least 20 eV. For a $\sim 9$ eV
multiplet, the excitation rate would be insensitive to this or higher
electron temperatures.  With the above assumptions at
$T_e \sim 20$ eV applied to the polar atmosphere on open field lines,
electron impact dissociative excitation of O$_2$ yields 300 R of
\Oone]\,\lam1356 at high latitudes with limb brightening to $\sim 1$ kR. A
polar limb intensity of 50 R implies an O$_2$ column density of $3 \times
10^{13}$ cm$^{-2}$.   Alternatively, in the absence of a direct measurement
of Jovian electron temperature, the upper limit column density of $5.2 \times
10^{14}$ cm$^{-2}$ is permissible, if $T_e \sim 4$ eV.  Also possible would
be various combinations of O$_2$ column density in the range of $(0.3-5)
\times 10^{14}$ cm$^{-2}$ and electron temperature in the range of 4--20 eV
or higher.

There are additional factors to consider in the interpretation of the
HST images.  There is no evidence that Ganymede's surface temperature
drops to the O$_2$ condensation temperature of 80 K which could account
for an inhomogeneous atmosphere. In the range $T_e = 1-100$ eV, the
calculated intensity ratio of \Oone]\,\lam1356 to \Oone\,\lam1304 is
limited to 1.6--2.0 for a pure O$_2$ atmosphere.  Lower values require
the addition of atomic oxygen.  In the limit of a pure atomic oxygen
atmosphere, this intensity ratio has a value of 1.2 at $T_e = 4$ eV,
monotonically decreasing to $\sim 0.35$ at 20 eV.  There is no apparent
correlation of this ratio in Table~\ref{tab1} with absolute
\Oone]\,\lam1356 brightness.  Clearly to achieve the higher observed
ratios requires an O$_2$ atmosphere.  Finally, Jovian magnetospheric
electrons on open field lines do not have large density
variability on the length scales characteristic of the brightness
variations in Fig.~\ref{oimage} \citep{gur96}.

Thus, the best explanation for the inhomogeneity in the emission
brightness is that it is truly auroral in nature, analogous with the
Earth's highly variable UV luminosity in the auroral oval regions and
driven by acceleration processes of electrons trapped within Ganymede's
magnetosphere near the separatrix regions.  Without knowledge of the
distribution function of these auroral precipitating electrons and
associated fluxes, the column density of Ganymede's atmosphere cannot be
inferred from the bright auroral regions in HST images.  For the present
time the polar limb intensities are the only constraint on the atmospheric
column density and until the temperature of electrons on open field lines
is determined, this constraint is not firm.

\subsection{\Hone\ \LA}

\begin{figure}
\begin{center}
\epsfig{file=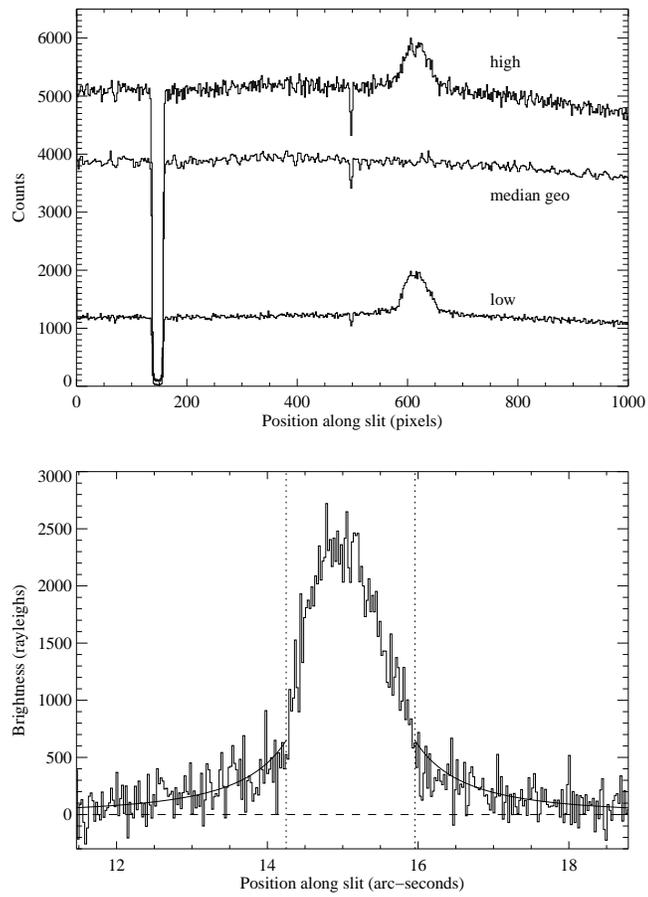, width=3.5in}
\figcaption{Top: Spatial profiles of \LA\ along the slit. ``High" and ``low"
refer to the first and second exposures, respectively, of orbits 2--4.
``Median geo" is the difference smoothed with a 5-point median filter.
Bottom: Spatial profile of \LA\ associated with Ganymede obtained by
subtracting the derived geocoronal background.  The model of
\cite{bar97} based on fitting Galileo UVS data is shown.
\label{lymana}}
\end{center}
\end{figure}

\citet{bar97} have reported the detection of \Hone\ \LA\ emission above
the limb of Ganymede extending nearly one Ganymede radius (2634 km),
which they attribute to a hydrogen exosphere.  Such emission should be
detectable in our long-slit spectral image but is masked by the strong
geocoronal \LA\ emission that fills the entire $2''$ wide slit (see
Figure~\ref{rawimage}).  To remove the geocoronal component, a \LA\ ``flat
field'' along the slit is needed.  This is obtained from our data in the
following manner.  The final three orbits contain separate spectral images
with distinctly different values of geocoronal background, 15 kR for the
first of each pair, 3.5 kR for the second.  The three ``high'' images and
the three ``low'' images are separately combined (again using the flat-fielded
counts files rather than the flux calibrated files), and spatial profiles
along the slit (summing 82 pixels in the dispersion direction) are obtained.
These are shown in the top panel of Figure~\ref{lymana}.  The profiles
are normalized to the slightly different cumulative exposure times and
the difference is taken, which eliminates the signal due to Ganymede, and
this is also shown (after median filtering) in the figure.  The geocoronal
background is then normalized to and subtracted from the ``low'' image
giving the net \LA\ spatial profile associated with Ganymede, as shown in
the lower panel of Figure~\ref{lymana}, where emission above both limbs is clearly detected.  The radial model of \citet{bar97}, integrated across
the width of the STIS slit, is also shown in the figure and is found to
fit our data very well.
\placefigure{lymana}

\section{Conclusions}

Objective grating images of Ganymede obtained with HST/STIS
show clearly separated \Oone\ emissions confirming the result of
\citet{hal98} that the emissions are confined to polar regions
(latitudes above 45\deg).  The total fluxes are consistent
with those reported by \citeauthor{hal98} but appear to vary in time
and in the relative intensities between northern and southern
hemispheres.  The \Oone]\,\lam1356/\Oone\ \lam1304 ratio is consistent
with the primary excitation mechanism being electron impact on O$_2$,
as postulated by \citeauthor{hal98}  While the spatial distribution of
the emissions is consistent with current models of the magnetic field
of Ganymede, expected longitudinal uniformity and limb brightening are
not observed.  In addition, \LA\ limb emission from a hydrogen
exosphere is detected and the measured brightness is found to be in
good agreement with the Galileo UVS observations of Barth
\etal\ (1997).

\acknowledgments

This work is based on observations with the National Aeronautics and
Space Administration -- European Space Agency HST obtained at the Space
Telescope Science Institute, which is operated by the Association of
Universities for Research in Astronomy, Incorporated, under NASA
contract NAS5-26555.  We acknowledge partial support by NASA contract
NAS5-30403 to the Johns Hopkins University.



\begin{table}
\begin{center}
\caption{Image Parameters and Derived Fluxes \label{tab1}}
\medskip
\begin{tabular}{ccccccc}
\tableline\tableline
Exposure & Start & Exposure & \Oone\ \lam1304 & \Oone]\,\lam1356 & Ratio & Projected \\
ID & Time (UT) & Time (s) & Flux\tablenotemark{a} & Flux\tablenotemark{a}
 & \lam1356/\lam1304 & Angle of $B_J$\tablenotemark{b} \\
\tableline
O53K01010  & 8:21:19 &   850.0  & $12.5 \pm 2.1$ & $19.5 \pm 2.3$ & $1.6 \pm 0.3$ & 17.1 \\
O53K01020  & 8:38:57 &   850.0  & $11.5 \pm 2.0$ & $25.1 \pm 2.3$ & $2.2 \pm 0.4$ & 15.8 \\[6pt]
O53K01030  & 9:40:09 &  1205.0  & $14.5 \pm 5.3$ & $22.5 \pm 2.2$ & $1.5 \pm 0.6$ & 7.8 \\
O53K01040  & 10:07:22 &  1205.0 & $10.4 \pm 1.8$ & $17.5 \pm 2.0$ & $1.7 \pm 0.3$ & 3.3 \\[6pt]
O53K01050  & 11:16:55 &  1125.0 & $11.6 \pm 5.7$ & $37.7 \pm 2.7$ & $3.2 \pm 1.6$ & --8.2 \\
O53K01060  & 11:42:48 & 1100.0 & $20.5 \pm 2.2$ & $30.4 \pm 2.4$ & $1.5 \pm 0.2$ & --11.6 \\[6pt]
O53K01070  & 12:53:42 & 1200.0 & $22.1 \pm 5.6$ & $26.1 \pm 2.6$ & $1.2 \pm 0.3$ & --17.0 \\
O53K01080  & 13:17:10 & 1130.0 & $16.5 \pm 2.3$ & $20.2 \pm 2.4$ & $1.2 \pm 0.2$ & --17.2 \\
\tableline
\end{tabular}
\end{center}
\tablenotetext{a}{$10^{-5} \phots$.  Quoted errors are $1\sigma$ statistical
uncertainty.}
\tablenotetext{b}{in degrees measured east of Jovian north (south pole).}
\end{table}




\end{document}